\documentclass[11pt,a4paper]{article}
\usepackage{comment}
\usepackage{graphicx}
\usepackage{amssymb}
\usepackage{amsmath}
\usepackage{amsfonts}
\usepackage{dsfont}
\usepackage{mathtools}
\usepackage{array}
\usepackage{soul}
\usepackage{rotating}
\usepackage{bbold,amsfonts}
\allowdisplaybreaks

\usepackage[utf8]{inputenc}
\usepackage{bm}
\usepackage{xcolor}
\usepackage{float}
\usepackage{braket}
\usepackage{placeins}
\usepackage[height=8.8in,width=6.45in]{geometry}
\usepackage[font=small,labelfont=bf]{caption}
\usepackage[hidelinks]{hyperref}
\usepackage{booktabs,float,slashed}
\usepackage{standalone}
\usepackage{caption}
\usepackage{subcaption}
\usepackage{makecell}
\usepackage[maxbibnames=99,sorting=none,giveninits=true,backend=biber,style=numeric-comp,sortcites,doi=false,hyperref=true]{biblatex}
\renewbibmacro{in:}{}
\usepackage{hyperref}
\DeclareFieldFormat{doilink}{\iffieldundef{doi}{#1}{\href{https://doi.org/\thefield{doi}}{#1}}}
\DeclareFieldFormat[article,periodical]{volume}{\mkbibbold{#1}}
\DeclareFieldFormat[article,periodical]{journaltitle}{#1}
\DeclareFieldFormat[article,periodical]{pages}{#1}

\usepackage{xpatch}
\xpatchbibdriver{article}
  {\usebibmacro{journal+issuetitle}%
   \newunit
   \usebibmacro{byeditor+others}%
   \newunit
   \usebibmacro{note+pages}}
  {\printtext[doilink]{%
     \usebibmacro{journal+issuetitle}%
     \newunit
     \usebibmacro{byeditor+others}%
     \newunit
     \usebibmacro{note+pages}}}
  {}{}

\numberwithin{equation}{section}

\newcommand{\beq}{\begin{equation}}
\newcommand{\eeq}{\end{equation}}

\makeatletter
\newcommand*{\letterdef@}{}
\newcommand*{\letterdef}[3]{%
	\def\letterdef@##1{\expandafter\newcommand\csname #1\endcsname{#2{##1}}}%
	\@tfor\@tempa :=#3\do{\expandafter\letterdef@\expandafter{\@tempa}}}
\makeatother
\letterdef{c#1} {\mathcal}{ABCDEFGHIJKLMNOPQRSTUVWXYZ} 
\letterdef{rm#1}{\mathrm} {dDeimM} 

\title{Integrated correlators with a Wilson line\\ in $\mathcal{N}=2$ SCFTs at strong coupling}

\begin{document}

\begin{titlepage}

\begin{flushright}
\small
\texttt{HU-EP-24/31}
\end{flushright}

\vspace*{10mm}
\begin{center}
{\LARGE \bf 
Integrated correlators with a Wilson line in 

\vspace{0.2cm}

a $\mathcal{N}=2$ quiver gauge theory  at strong coupling 
}

\vspace*{15mm}

{\Large  Alessandro Pini${}^{\,a}$}

\vspace*{8mm}

${}^a$ Institut f{\"u}r Physik, Humboldt-Universit{\"a}t zu Berlin,\\
     IRIS Geb{\"a}ude, Zum Großen Windkanal 2, 12489 Berlin, Germany  
     \vskip 0.3cm

\vskip 0.8cm
	{\small
		E-mail:
		\texttt{alessandro.pini@physik.hu-berlin.de}
	}
\vspace*{0.8cm}
\end{center}

\begin{abstract}
We consider the four-dimensional $\mathcal{N}=2$ quiver gauge theory arising from a $\mathbb{Z}_2$ orbifold of $\mathcal{N}=4$ super Yang-Mills with gauge group $SU(2N)$. We study the integrated correlator between a half-BPS Wilson line and two Higgs branch operators of conformal dimension 2. Using supersymmetric localization, we resum the perturbative series for this observable and obtain exact expressions for both the planar and next-to-planar terms of its large $N$ expansion, valid for arbitrary values of the 't Hooft coupling. Finally, through a combination of analytical and numerical methods, we determine the leading terms of the corresponding strong-coupling expansions.
\end{abstract}
\end{titlepage}
\setcounter{tocdepth}{2}
\tableofcontents

\section{Introduction}
Four dimensional conformal field theories with extended supersymmetry (SCFTs) have been extensively studied in recent years, as they provide an ideal framework for obtaining exact results and exploring non-perturbative physics. 

In particular, increasing attention has been devoted to the study of integrated correlators of local operators in $\mathcal{N}=4$ super Yang-Mills (SYM), the maximally supersymmetric theory in four dimensions. These observables correspond to 4-point correlation functions of scalar operators, integrated with a specific measure fully determined by superconformal symmetry. They can be computed exactly through supersymmetric localization \cite{Pestun:2007rz,Pestun:2016jze}, by placing the theory on a unit 4-sphere $\mathbb{S}^4$ and considering the mass deformation of $\mathcal{N}=4$ SYM, known as $\mathcal{N}=2^{*}$,
in which the $\mathcal{N}=2$ adjoint hypermultiplet acquires a mass $m$. Notably, it has been shown that taking four derivatives of the logarithm of the mass-deformed partition function $\mathcal{Z}_{\mathcal{N}=2^{*}}$ yields \cite{Binder:2019jwn,Chester:2020dja}
\begin{subequations}
\begin{align}
& \partial_{\tau_p}\partial_{\bar{\tau}_p}\partial_{m}^2\log \mathcal{Z}_{\mathcal{N}=2^{*}}\Big|_{m=0} = \int \prod_{i=1}^{4}dx_i\, \mu(x_i)\, \langle \mathcal{O}_p(x_1)\mathcal{O}_p(x_2)\mathcal{O}_2(x_3)\mathcal{O}_2(x_4) \rangle_{\mathcal{N}=4}\\
& \partial_{m}^4\log \mathcal{Z}_{\mathcal{N}=2^{*}}\Big|_{m=0} = \int \prod_{i=1}^{4}dx_i\, \mu'(x_i)\, \langle \mathcal{O}_2(x_1)\mathcal{O}_2(x_2)\mathcal{O}_2(x_3)\mathcal{O}_2(x_4) \rangle_{\mathcal{N}=4}\, \ ,
\end{align}
\label{IntegratedFirstType}
\end{subequations}
where $\mathcal{O}_p(x_i)$ denotes a half-BPS chiral scalar operator with conformal dimension $p$, transforming in the $[0,p,0]$ representation of the $SU(4)_R$ $R$-symmetry group. The couplings $\tau_p$ and $\bar{\tau}_p$ are associated with the chiral and anti-chiral operators $\mathcal{O}_p(x_i)$, respectively \cite{Gerchkovitz:2016gxx}. Finally, $\mu(x_i)$ and $\mu'(x_i)$ represent the two integration measures. In recent years, many properties of these observables have been thoroughly investigated, with particular focus on the modular properties and large-$N$ expansion using topological recursion \cite{Chester:2019jas,Dorigoni:2021bvj,Dorigoni:2021guq,Chester:2020vyz,Collier:2022emf,Dorigoni:2022cua,Paul:2022piq,Dorigoni:2024dhy,Alday:2023pet}, the extension to all simple gauge groups \cite{Dorigoni:2022zcr,Dorigoni:2023ezg}, the weak-coupling expansion and the check of the localization predictions through explicit Feynman diagram computations \cite{Wen:2022oky,Zhang:2024ypu}, the generalization to cases involving operators with arbitrary conformal dimensions \cite{Brown:2023zbr} and the insertion of determinant operators \cite{Brown:2024tru}, as well as their large-charge limit \cite{Paul:2023rka, Brown:2023cpz, Brown:2023why}. Finally many features of the integrated correlators \eqref{IntegratedFirstType} have also been investigated in the context of purely $\mathcal{N}=2$ SCFTs \cite{Chester:2022sqb, Fiol:2023cml, Behan:2023fqq, Billo:2023kak, Pini:2024uia, Billo:2024ftq}.

However, comparatively less attention has been devoted to the study of a different class of integrated correlators—those involving two chiral operators of conformal dimension 2 and a line defect, such as a half-BPS Wilson line. This observable can still be computed  using supersymmetric localization by taking two derivatives of the logarithm of the Wilson line vacuum expectation value $\langle W \rangle$ in the mass-deformed $\mathcal{N}=2^{*}$ theory \cite{Pufu:2023vwo}
\begin{align}
\label{Wint}
\mathcal{I} \equiv \partial_{m}^2 \log \langle W \rangle \big|_{m=0} = \int d^4x_1 \, d^4x_2 \, \hat{\mu}(x_1, x_2) \, \langle \mathcal{O}_2(x_1) \mathcal{O}_2(x_2) \rangle_{W} \, ,
\end{align}
where $\langle \mathcal{O}_2(x_1) \mathcal{O}_2(x_2) \rangle_W$ denotes the 2-point function between the operators $\mathcal{O}_2(x_i)$ in the presence of a Wilson line. Furthermore, the explicit expression of the integration measure $\hat{\mu}(x_1, x_2)$, which is fully determined by superconformal symmetry, has been obtained in \cite{Billo:2023ncz,Dempsey:2024vkf,Billo:2024kri}. From a mathematical perspective, the relation \eqref{Wint} imposes an integral constraint on local operators in the presence of a Wilson line. Together with the constraints from \eqref{IntegratedFirstType}, it can be utilized in subsequent numerical bootstrap computations, enabling the derivation of numerical bounds on anomalous dimensions and OPE coefficients (see e.g. \cite{Chester:2022sqb,Chester:2023ehi}). However, the importance of the second type of integrated correlator \eqref{Wint} also lies in the fact that, in general, defect operators transform non-trivially under the $\mathcal{S}$-duality group \cite{MONTONEN1977117,WITTEN197897} of $\mathcal{N}=4$ SYM, while the local operators $\mathcal{O}_p(x)$ remain invariant. By exploiting this feature, the modular properties of the integrated correlator \eqref{Wint}, along with the case involving the insertion of a generic dyonic half-BPS defect operator, have been examined in \cite{Pufu:2023vwo, Dorigoni:2024vrb, Dorigoni:2024csx}.

As discussed in \cite{Billo:2023ncz,Dempsey:2024vkf,Billo:2024kri}, the relation \eqref{Wint} also holds in the context of purely $\mathcal{N}=2$ SCFTs. In this case, however, the operators $\mathcal{O}_2(x_i)$ are replaced by the so-called ``moment map'' operators, i.e. Higgs branch operators with  conformal dimension equal to 2. These operators arise as the top component of the short multiplet $\mathcal{\hat{B}}_1$ of the $\mathfrak{su}(2,2|2)$ superconformal algebra\footnote{We use the conventions of \cite{Dolan:2002zh} for denoting the multiplets of the $\mathfrak{su}(2,2|2)$ algebra.} and are constructed as bilinears of the scalar fields inside the $\mathcal{N}=2$ hypermultiplets.

Despite their importance and the progress made in the maximally supersymmetric theory, to the best of our knowledge, explicit evaluations of the integrated correlator \eqref{Wint} in a purely $\mathcal{N}=2$ SCFT are still lacking. In this article, we intend to take the first step in this direction and aim to address this gap by focusing on the $\mathcal{N}=2$ two-node quiver gauge theory that arises as an orbifold of $\mathcal{N}=4$ SYM with gauge group $SU(2N)$. Specifically, we concentrate on the so-called ``orbifold fixed point" of the theory, defined by the condition that the two Yang-Mills coupling constants associated with the gauge groups are equal. This choice is motivated by the fact that, for this symmetric configuration, the theory has a known gravitational dual geometry described by type II B string theory on $\text{AdS}_5 \times \mathbb{S}^5 / \mathbb{Z}_2$ \cite{Kachru:1998ys,Gukov:1998kk,Skrzypek:2023fkr}. Consequently, the results we obtain in the 't Hooft limit of the theory, can furnish predictions for a subsequent holographic analysis. Furthermore we can still employ supersymmetric localization, which allows us to reformulate the evaluation of the Wilson loop expectation value in the mass-deformed theory as the computation of finite-dimensional integrals, that can be calculated through a matrix model. It is worth noting that for $\mathcal{N}=4$ SYM, the matrix model obtained via localization is Gaussian, a feature that facilitates obtaining results both at finite $N$ and in the large-$N$ limit. On the other hand, for a generic $\mathcal{N}=2$ theory, the matrix model involves a non-trivial potential, making it challenging to derive results, particularly in the strong coupling regime. Nevertheless, over the past few years, it has become evident that the specific structure of the interaction of the aforementioned quiver gauge theory enables the exact computation of various observables in the large-$N$ limit. These include the free energy and correlators among half-BPS Wilson loops \cite{Rey:2010ry,Zarembo:2020tpf,Fiol:2020ojn,Ouyang:2020hwd,Beccaria:2021ksw,Galvagno:2021bbj,Pini:2023lyo,Beccaria:2023kbl} and 2- and 3-point correlation functions among chiral/anti-chiral scalar operators \cite{Pini:2017ouj,Galvagno:2020cgq,Fiol:2021icm,Fiol:2022vvv,Billo:2021rdb,Billo:2022gmq,Billo:2022fnb,Billo:2022lrv}. Consequently we find it natural to investigate whether this feature can also be used to study the integrated correlator \eqref{Wint}.

The rest of this article is organized as follows. In Section \ref{sec:Z2QuiverTheory}, we review the main features of the $\mathcal{N}=2$ quiver gauge theory and introduce the matrix model representation of the mass deformed partition function, that constitutes the starting point of our analysis. We then recall the key properties of this theory in the large-$N$ limit and review how computations become more manageable in this regime. Then, in Section \ref{sec:LargeNIntegartedCorrelator}, we derive exact expressions for the planar and next-to-planar terms of the integrated correlator \eqref{Wint}, valid for any value of the 't Hooft coupling $\lambda$. Finally, by exploiting the asymptotic properties of Bessel functions, we determine the first terms of the corresponding large $\lambda$ expansions. Our conclusions are presented in Section \ref{sec:Conclusions}, where we also discuss the potential generalization of our results to a generic $\mathbb{Z}_{M>2}$ quiver gauge theory. Furthermore, technical details regarding the mathematical proof of the identities used in deriving our results, as well as a brief review of the techniques employed to perform the strong coupling expansions, are included in two appendices.

\section{The \texorpdfstring{ $\mathbb{Z}_2$}{} quiver gauge theory}
\label{sec:Z2QuiverTheory}
\begin{figure}[ht]
    \centering
\includegraphics[scale=1.0]{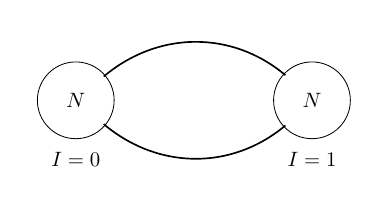}
\caption{The $4d$ $\mathcal{N}=2$ quiver gauge theory.}
    \label{fig:quiverZ2}
\end{figure}

We consider the $4d$ $\mathcal{N}=2$ quiver gauge theory arising as a $\mathbb{Z}_2$ orbifold of $\mathcal{N}=4$ SYM with gauge group $SU(2N)$. We summarize its matter content with the quiver diagram reported in Figure \ref{fig:quiverZ2}. Each node (labelled by an index $I=0,1$) represents an $SU(N)$ gauge group, and each line corresponds to an $\mathcal{N}=2$ hypermultiplet in the bifundamental representation. Furthermore, we focus only on the so-called ``orbifold fixed point", where the two coupling constants $g_0$ and $g_1$ are set equal to a common value $g$. This allows us to define a unique 't Hooft coupling, namely $\lambda \equiv g^2 N$. Finally, we observe that the theory is conformal, as the $\beta$-function vanishes for both gauge nodes. 

The gauge invariant defect operator on which we focus our attention is the half-BPS circular Wilson loop  in the fundamental representation of the $I$-th node of the quiver on a 4-sphere with radius $\mathcal{R}$, namely
\begin{align}
\label{WilsonLoop}
W^{(I)} = \frac{1}{N}\,\text{tr}\,\mathcal{P}\,\exp\,\Bigg\{g\oint_{\mathcal{C}}d\tau \left[i\,A_{\mu}^{I}(x)\,\dot{x}^{\mu}(\tau)+\frac{\mathcal{R}}{2}\left(\varphi^{I}(x)+\overline{\varphi}^{I}(x)\right)\right]\Bigg\}\, \,  ,   
\end{align}
where $\mathcal{C}$ denotes the equator of the 4-sphere, $A_{\mu}^{I}(x)$ and $\varphi^{I}(x)$ represent the vector field and the complex scalar field of the $I$-th $\mathcal{N}=2$ vector multiplet. Finally $\mathcal{P}$ denotes the path ordering. Without loss of generality, we choose the coordinates in such a way that the equator $\mathcal{C}$ lies in the $\mathbb{R}^2 \subset \mathbb{R}^4$ parametrized by $(x_1, x_2)$. Thus, the function $x^{\mu}(\tau)$ can be explicitly expressed as follows
\begin{align}
x^{\mu}(\tau) = \mathcal{R}\,(\cos(\tau),\sin(\tau),0,0)\, \, \,  \qquad \qquad 0 \leq \tau < 2\pi\, \ .    
\end{align}
Now on we consider a unit 4-sphere, i.e. we set $\mathcal{R} = 1$.
As discussed in \cite{Pini:2023lyo}, it is useful to consider two distinct linear combinations, namely
\begin{align}
\label{Wut}
W_{\pm} = \frac{1}{\sqrt{2}}\left(W^{(0)}\pm W^{(1)}\right)\, \ .
\end{align}
The operators \(W_{+}\) and \(W_{-}\) are referred to as the untwisted $(+)$ and twisted $(-)$ Wilson loop respectively, since they are even and odd under the action of the \(\mathbb{Z}_2\) symmetry that exchanges the two nodes of the quiver diagram of Figure \ref{fig:quiverZ2}.

In the next subsection, we explain how the computation of the  vacuum expectation value of \eqref{Wut}, as well as that of the integrated correlator \eqref{Wint}, can be performed using an interacting matrix model.

\subsection{The mass deformed matrix model}
Henceforth we consider the mass-deformed theory obtained by giving a mass $m_1$ to the first $\mathcal{N}=2$ hypermultiplet and a mass $m_2$ to the second. By placing it on a unit 4-sphere and applying supersymmetric localization \cite{Pestun:2007rz}, we can rewrite the partition function of this theory in the following form
\begin{align}
\label{Zmassive}
\mathcal{Z}(m_1,m_2) =  \left(\frac{8\pi^2\,N}{\lambda}\right)^{N^2-1}\int da \int db\, \text{e}^{-\frac{8\pi^2N}{\lambda}(\textrm{tr}a^2+\textrm{tr}b^2)}\Big|\mathcal{Z}_{\textrm{1-loop}}(m_1,m_2,a,b)\mathcal{Z}_{\textrm{inst}}(a,b,m_1,m_2,\lambda)\Big|^2\, \ ,    
\end{align}
where we employ the so-called ``full Lie algebra approach" (introduced in \cite{Billo:2017glv}), and therefore the integration is performed over all elements of the two Hermitian traceless matrices $a$ and $b$, which are associated with the two gauge nodes.\footnote{The matrices $a$ and $b$ and the integration measures $da$ and $db$ are explicitly given by
\begin{align}
a = \sum_{c=1}^{N^2-1} a_c T^c,\, \, \, \, b = \sum_{d=1}^{N^2-1} b_d T^d, \qquad \qquad da = \prod_{c=1}^{N^2-1} \frac{da_c}{\sqrt{2\pi}}\, , \,\, \, \, db = \prod_{d=1}^{N^2-1} \frac{db_d}{\sqrt{2\pi}}\, \ ,
\end{align}
where $T_c$ are the generators of the $\mathfrak{su}(N)$ Lie algebra, normalized such that $\text{tr}(T_e T_f) = \frac{1}{2}\delta_{e,f}$.
} Moreover $\mathcal{Z}_{1\text{-loop}}$ and $\mathcal{Z}_{\text{inst}}$ represent the 1-loop and instanton contributions, respectively. Since in this work we focus only on the large $N$-limit of the theory with fixed $\lambda$, where the instanton contribution is exponentially suppressed, we will henceforth set $\mathcal{Z}_{\text{inst}} = 1$.

To compute the integrated correlator \eqref{Wint} we need to consider the small mass expansion of the partition function \eqref{Zmassive} up to order $O(m^2)$, which has already been performed in  \cite{Pini:2024uia}. Therefore, here we just report the result, which is as follows
\begin{align}
\label{Zexpansion}
\mathcal{Z}(m_1,m_2) = \int da \int db\, \text{e}^{-(\textrm{tr}\,a^2+\textrm{tr}\,b^2)}\,\textrm{e}^{-S_0+(m_1^2+m_2^2)\mathcal{M}_{\mathbb{Z}_2}+o(m^2)}\, \ , 
\end{align}
with
\begin{align}
& S_0 = \sum_{m=2}^{\infty}\sum_{k=2}^{2m}(-1)^{m+k}\left(\frac{\lambda}{8\pi^2N}\right)^{m}\left(\begin{array}{c}2m\\
 k\end{array}\right)\frac{\zeta_{2m-1}}{m}\left(\text{tr}\,a^{2m-k}-\text{tr}\,b^{2m-k}\right)\left(\text{tr}\,a^k-\text{tr}\,b^k\right)\,, \label{S0Expansion}\\[0.5em]
& \mathcal{M}_{\mathbb{Z}_2} =-\sum_{n=1}^{\infty}\sum_{\ell=0}^{2n}(-1)^{n+\ell}\,\frac{(2n+1)!\,\zeta_{2n+1}}{(2n-\ell)!\,\ell!}\left(\frac{\lambda}{8\pi^2N}\right)^n\text{tr}\,a^\ell\,\text{tr}\,b^{2n-\ell} \, , \label{MZ2}
\end{align}
where $\zeta_{n}$ denotes the Riemann zeta function $\zeta(n)$. We observe that we can regard the term $S_0-(m_1^2+m_2^2)\mathcal{M}_{\mathbb{Z}_2}$ present in \eqref{Zexpansion} as an interaction action added to the Gaussian matrix model.

As shown, for instance in \cite{Pestun:2007rz,Pini:2023lyo}, the half-BPS circular untwisted/twisted Wilson loop \eqref{Wut} admits the following matrix model representation
\begin{align}
\label{WilsonMatrixModel}
\textsf{W}_{\pm}(a,b,\lambda) = \frac{1}{\sqrt{2}\,N}\left[\text{tr} \exp\Bigg\{\sqrt{\frac{\lambda}{2N}}\,a\Bigg\}\pm \text{tr}\exp\Bigg\{\sqrt{\frac{\lambda}{2N}}\,b\Bigg\} \right] = \frac{1}{N}\sum_{k=0}^{\infty}\frac{1}{k!}\left(\frac{\lambda}{2N}\right)^{\frac{k}{2}}A_k^{\pm}\, \ ,
\end{align}
where we introduced the untwisted (+) and twisted (-) combinations 
\begin{align}
\label{AA}
A_{k}^{\pm} = \frac{1}{\sqrt{2}}(\text{tr}\,a^k \pm \text{tr}\,b^k)\, .  
\end{align}
Therefore, the matrix model representation of the vacuum expectation value of \eqref{Wut} in the mass deformed theory, reads
\begin{align}
\label{Wvevmassive}
& \mathcal{W}_{\pm}(m_1,m_2) \equiv \nonumber \\
& \frac{1}{\mathcal{Z}(m_1,m_2)}\left(\frac{8\pi^2N}{\lambda}\right)^{N^2-1}\int da\int db \, \textsf{W}_{\pm}(a,b,\lambda) \, \text{e}^{-\frac{8\pi^2N}{\lambda}(\text{tr}\,a^2 + \text{tr}\,b^2)}\text{e}^{-S_0+(m_1^2+m_2^2)\mathcal{M}_{\mathbb{Z}_2} + o(m^2)}\, \ .  
\end{align}
Then, using the definition  \eqref{Wut}, it is straightforward to see that the vacuum expectation value of a twisted Wilson loop vanishes, i.e.  $\mathcal{W}_{-}(m_1,m_2) = 0$. Therefore, from this point onward, we will focus exclusively on the untwisted Wilson loop. Following \cite{Billo:2024kri,Dempsey:2024vkf} and by employing both \eqref{Wvevmassive} and \eqref{Zmassive}, we finally obtain the matrix model representation of the integrated correlator \eqref{Wint} for the $\mathbb{Z}_2$ quiver gauge theory, namely
\begin{align}
\label{IZ2}
\mathcal{I}_{\mathbb{Z}_2} \equiv \partial_{m_f}^2\log \mathcal{W}_{+}(m_1,m_2)\Big|_{m_1=m_2=0} =  2\,\frac{\langle \textsf{W}_{+}(a,b,\lambda)\,\mathcal{M}_{\mathbb{Z}_2} \rangle - \langle \textsf{W}_{+}(a,b,\lambda)\rangle \,\langle \mathcal{M}_{\mathbb{Z}_2} \rangle}{\langle \textsf{W}_{+}(a,b,\lambda) \rangle}         
\end{align}
with $f=1,2$ and where the vacuum expectation value $\langle \cdot \rangle$ has been computed with respect to the massless matrix model. For a generic operator $\mathcal{O}(a,b)$ this expectation value is defined as
\begin{align}
\langle \mathcal{O}(a,b) \rangle = \frac{\langle \mathcal{O}(a,b)\,\text{e}^{-S_0} \rangle_0}{\langle \text{e}^{-S_0} \rangle_0}\, \ ,  
\end{align}
where $\langle \cdot \rangle_0$ stands for the v.e.v. in the free matrix model. This, in turn, can be efficiently evaluated by exploiting the set of identities satisfied by the functions \cite{Beccaria:2020hgy, Billo:2017glv}
\begin{align}
t_{n_1,\ldots,n_{M}} \equiv \langle \text{tr}\,a^{n_1} \cdots \, \text{tr}\,a^{n_M} \rangle_0\, ,
\end{align}
where $a$ denotes a generic matrix in the $\mathfrak{su}(N)$ Lie algebra. A brief inspection of the expressions \eqref{MZ2} and \eqref{WilsonMatrixModel} clearly shows that evaluating \eqref{IZ2} involves computing correlators of the form \(\langle A^{\pm}_{k_1}\,\cdots\,A^{\pm}_{k_M} \rangle\). Therefore, in the next section, we will review the mathematical tools that enable us to efficiently perform such computations in the large \(N\) limit of the \(\mathbb{Z}_2\) quiver gauge theory.

\subsection{Properties of the \texorpdfstring{$\mathbb{Z}_2$}{} quiver gauge theory at large \texorpdfstring{$N$}{}}
Although the operators \eqref{AA} naturally appear in the expressions for the untwisted and twisted Wilson loops \eqref{WilsonMatrixModel}, it was demonstrated in \cite{Billo:2022fnb} that computations in the large-$N$ limit of the \(\mathbb{Z}_2\) quiver gauge theory can be performed more efficiently by employing a different operator basis, denoted as $\mathcal{P}_k^{\pm}$ and defined by the following relation
\begin{align}
\label{PP}
A_{k}^{\pm} = \left(\frac{N}{2}\right)^{\frac{k}{2}}\sum_{\ell=0}^{\lfloor\frac{k-1}{2}\rfloor}\sqrt{k-2\ell}\left(\begin{array}{c}
     k  \\
     \ell 
\end{array}\right)\mathcal{P}_{k-2\ell}^{\pm}   + \langle A_{k}^{\pm} \rangle_0 \, .
\end{align}
One of the benefits of this choice is that, in the free theory and in the large-$N$ limit, the operators $\mathcal{P}_{k}^{\pm}$ are orthonormal, namely\footnote{From this point onward, the symbol $\simeq$ will denote equality at the leading order of the large-$N$ expansion.}
\begin{align}
\langle \mathcal{P}_{k}^{\pm}\,\mathcal{P}_{\ell}^{\pm} \rangle_0 \, \simeq \, \delta_{k,\ell} \, .
\end{align}
Moreover, the interaction action  \eqref{S0Expansion} of the massless theory has a very simple form in the basis of the  $\mathcal{P}_{k}^{-}$ operators \cite{Billo:2021rdb,Billo:2022fnb}
\begin{align}
S_0 \,=\, -\frac{1}{2}\sum_{k,\ell=1}^{\infty}\mathcal{P}_{k}^{-}\,\textsf{X}_{k,\ell}\,\mathcal{P}_{\ell}^{-}\, \ ,
\label{S0viaX}
\end{align}
where the $\textsf{X}$-matrix is given by
\begin{align}
\textsf{X}_{k,\ell} \,=\, 2\,(-1)^{\frac{k+\ell+2k\ell}{2}+1}\sqrt{k\,\ell}\int_0^{\infty} \frac{dt}{t}\,\frac{1}{\sinh(t/2)^2}\,J_k\left(\frac{t\sqrt{\lambda}}{2\pi}\right)J_{\ell}\left(\frac{t\sqrt{\lambda}}{2\pi}\right)
\label{Xmatrix}
\end{align}
for $ k, \ell \geq 2$, and the entries with opposite parities vanish, i.e. $\textsf{X}_{2k, 2\ell + 1} = 0$. It is worth noting that, although $S_0$ was defined in \eqref{S0Expansion} as a series in powers of $\lambda$, the expression \eqref{Xmatrix} is exact in the 't Hooft coupling, as all the dependence has been summed in terms of Bessel functions of the first kind. Therefore, it turns to be particularly useful for studying the strong coupling regime of the theory as it was done for example in \cite{Billo:2022lrv,Pini:2024uia}.
Furthermore, the relation \eqref{S0viaX} allows to express the partition function $\mathcal{Z}$ and the free energy $\mathcal{F}$ of the massless theory in terms of the $\textsf{X}$-matrix as follows \cite{Billo:2021rdb}
\begin{align}
\mathcal{Z} = \left(\text{det}[\mathbb{1}-\textsf{X}]\right)^{-\frac{1}{2}}\, ,\qquad \qquad \mathcal{F}=\frac{1}{2}\text{tr}\log(\mathbb{1}-\textsf{X})\, \ .   
\end{align}
It then follows that, in the large-$N$ limit,  the 1-, 2- and 3-point correlation functions among $\mathcal{P}_{k}^{\pm}$ operators can also be expressed in terms of the $\textsf{X}$-matrix and its powers. In particular, as shown in \cite{Billo:2022fnb,Pini:2024uia}, it holds that 
\begin{subequations}
\begin{align}
& \langle \mathcal{P}_{k}^{-} \rangle = \langle \mathcal{P}_{2k+1}^{+} \rangle =0\, , \, \langle \mathcal{P}_{2k}^{+} \rangle \simeq -\frac{\sqrt{k}}{N}\lambda\partial_{\lambda}\mathcal{F}\, \ ,
\label{1pt}\\[0.3em]
& \langle \mathcal{P}_{k}^{-}\,\mathcal{P}_{\ell}^{-} \rangle \simeq \textsf{D}_{k,\ell}\ , 
\label{2pt}\\[0.0em]
& \langle \mathcal{P}_{2k}^{+}\,\mathcal{P}_{2\ell}^{+} \rangle = \delta_{k,\ell} \, + \, \frac{\textsf{T}^{+}_{k,\ell}}{N^2} + O(N^{-4})\, \ , \\[0.3em]
& \langle \mathcal{P}_{k_1}^{+}\,\mathcal{P}_{k_2}^{-}\,\mathcal{P}_{k_3}^{-} \rangle \, \simeq \, \frac{\sqrt{k_1}\,\textsf{d}_{k_2}\,\textsf{d}_{k_3}}{\sqrt{2}N} +\langle \mathcal{P}_{k_2}^{-}\,\mathcal{P}_{k_3}^{-} \rangle \, \langle \mathcal{P}_{k_1}^{+} \rangle \,  ,\\
& \langle \mathcal{P}^{+}_{k_1}\,\mathcal{P}^{+}_{k_2}\,\mathcal{P}^{+}_{k_3} \rangle \simeq \frac{\sqrt{k_1k_2k_3}}{\sqrt{2}\,N} + \langle \mathcal{P}^{+}_{k_1} \rangle\delta_{k_2,k_3} +  \langle \mathcal{P}^{+}_{k_2} \rangle\delta_{k_1,k_3}+  \langle \mathcal{P}^{+}_{k_3} \rangle\delta_{k_1,k_2}\, \ ,
\end{align}
\label{PPcorrelators}
\end{subequations}
with $k_1+k_2+k_3$ being even, and where
\begin{align}
\textsf{D}_{k,\ell} = \delta_{k,\ell} + \textsf{X}_{k,\ell} + (\textsf{X}^2)_{k,\ell} + (\textsf{X}^3)_{k,\ell} + \dots\, \ , \qquad \qquad \textsf{d}_k = \sum_{\ell=2}^{\infty}\sqrt{\ell}\,\textsf{D}_{k,\ell} \, ,
\label{dandD}
\end{align}
and
\begin{align}
\textsf{T}^{+}_{k,\ell} = \sqrt{k\ell}\left[\frac{(k^2+\ell^2-1)(k^2+\ell^2-14)}{12}-(k^2+\ell^2-1)\lambda\partial_{\lambda}\mathcal{F} - (\lambda\partial_{\lambda})^2\mathcal{F}+(\lambda\partial_{\lambda}\mathcal{F})^2\right]\, \ .   
\end{align}
Finally, we note that also the operator \eqref{MZ2} can be decomposed on the $\mathcal{P}_{k}^{\pm}$ basis, namely
\begin{align}
\mathcal{M}_{\mathbb{Z}_2} \,=\, \mathcal{M}_{\mathbb{Z}_2}^{(0)} + \mathcal{M}_{\mathbb{Z}_2}^{(1)}+
\mathcal{M}_{\mathbb{Z}_2}^{(2)}\, ,
\label{Mexpansion}
\end{align}
where the three terms on the r.h.s. contain zero, one and two $\mathcal{P}_{k}^{\pm}$ operators, respectively. In the large $N$ limit they admit the following expansions \cite{Pini:2024uia} 
\begin{subequations}
\label{M2full}
\begin{align}
& \mathcal{M}_{\mathbb{Z}_2}^{(0)} = N^2\,\mathsf{M}_{0,0} + \mathsf{M}_{1,1} -\frac{1}{6}\sum_{k=1}^{\infty}\sqrt{2k+1}\mathsf{M}_{1,2k+1} + O\left(N^{-2}\right) \,, \\
& \mathcal{M}_{\mathbb{Z}_2}^{(1)} = \sqrt{2}N\sum_{k=1}^{\infty}\mathsf{M}_{0,2k}\mathcal{P}_{2k}^{+} + \frac{\sqrt{2}\,\lambda}{32\pi^2 N}\sum_{k=1}^{\infty}\mathsf{Q}_{0,2k}\mathcal{P}_{2k}^{+} - \frac{\lambda}{192\pi^2 N}\sum_{k=1}^{\infty}\mathsf{Q}_{2,2k}\,\mathcal{P}_{2k}^{+} +O(N^{-3})\,, \label{M21} \\
& \mathcal{M}_{\mathbb{Z}_2}^{(2)} = \frac{1}{2}\sum_{p,q=2}^{\infty}(-1)^{p-pq}\,\mathsf{M}_{p,q}\,(\mathcal{P}_{p}^{+}\,\mathcal{P}_{q}^{+}-\mathcal{P}_{p}^{-}\,\mathcal{P}_{q}^{-})\, ,   \end{align}
\end{subequations}
and the coefficients $\mathsf{M}_{n,m}$ are defined as follows
\begin{subequations}
\begin{align}
\mathsf{M}_{0,0}&=\int_0^\infty\!\frac{dt}{t}\,\frac{(t/2)^2}{\sinh(t/2)^2} \bigg[1-\frac{16\,\pi^2}{t^2\lambda}\,J_1\Big(\frac{t\sqrt{\lambda}}{2\pi}\Big)^2\bigg]~,\label{M00}\\[2mm]
\mathsf{M}_{0,n}&=(-1)^{\frac{n}{2}+1}\,\sqrt{n}\!\int_0^\infty\!\frac{dt}{t}\,\frac{(t/2)^2}{\sinh(t/2)^2}\, \Big(\frac{4\pi}{t\sqrt{\lambda}}\Big)\,J_1\Big(\frac{t\sqrt{\lambda}}{2\pi}\Big)\,J_n\Big(\frac{t\sqrt{\lambda}}{2\pi}\Big)~,\label{M0n}\\[2mm]
\mathsf{M}_{n,m}&=(-1)^{\frac{n+m+2nm}{2}+1}\,\sqrt{nm}\!\int_0^\infty\!\frac{dt}{t}\,\frac{(t/2)^2}{\sinh(t/2)^2}
\,J_n\Big(\frac{t\sqrt{\lambda}}{2\pi}\Big)\,J_m\Big(\frac{t\sqrt{\lambda}}{2\pi}\Big)\label{Mnm}\,  .
\end{align}
\label{Mmatrix}%
\end{subequations}
While the coefficients $\textsf{Q}_{0,n}$ and $\textsf{Q}_{n,m}$, which appear in \eqref{M21} in the next-to-planar term of $\mathcal{M}_{\mathbb{Z}_2}^{(1)}$, read
\begin{subequations}
\begin{align}
& \textsf{Q}_{0,n} = (-1)^{\frac{n}{2}+1}\sqrt{n}\int_0^{\infty}\,dt\, t\, \frac{(t/2)^2}{\sinh(t/2)^2}\left(\frac{4\pi}{t\sqrt{\lambda}}\right)J_1\left(\frac{t\sqrt{\lambda}}{2\pi}\right)J_n\left(\frac{t\sqrt{\lambda}}{2\pi}\right)\, , \label{Q0n} \\[0.75em]
& \textsf{Q}_{n,m} = (-1)^{\frac{n+m+2nm}{2}+1}\sqrt{n\,m}\int_{0}^{\infty}\, dt\, t\, \frac{(t/2)^2}{\sinh(t/2)^2}J_n\left(\frac{t\sqrt{\lambda}}{2\pi}\right)J_m\left(\frac{t\sqrt{\lambda}}{2\pi}\right)\,  . \label{Qnm}
\end{align}
\label{Qmatrix}
\end{subequations}

\section{The large-\texorpdfstring{$N$}{} expansion of the integrated correlator \texorpdfstring{$\mathcal{I}_{\mathbb{Z}_2}$}{}}
\label{sec:LargeNIntegartedCorrelator}
In this Section we compute the first orders of the large-$N$ expansion of the integrated correlator \eqref{IZ2}. Specifically, we first analyze separately the vacuum expectation value of the untwisted Wilson loop and of the connected correlator appearing in the numerator of \eqref{IZ2}, then we evaluate the planar and next-to-planar terms of $\mathcal{I}_{\mathbb{Z}_2}$.

\subsection{The untwisted Wilson loop \texorpdfstring{$W_{+}$}{}}
Let's start by evaluating the large $N$ limit of the vacuum expectation value of the untwisted Wilson loop  in the massless interacting theory.  We consider the matrix model representation of this observable, given by the expression \eqref{WilsonMatrixModel}, and then perform the change of basis \eqref{PP}. In this way, we obtain
\begin{align}
\label{WuPbasis}
W_{+} = \frac{1}{N}\sum_{k=0}^{\infty}\frac{1}{k!}\left(\frac{\lambda}{2N}\right)^{\frac{k}{2}}A_{k}^{+} = \langle W_{+} \rangle_0 + \frac{1}{N}\sum_{q=2}^{\infty}\sqrt{q}\,I_q(\sqrt{\lambda})\,\mathcal{P}^{+}_q\, \ ,    
\end{align}
where $I_q(x)$ denotes the modified Bessel function of the first kind. Then, the large $N$ expansion of its vacuum expectation value up to the next to planar term, can be written as follows
\begin{align}
\label{vevW0}
\langle W_{+} \rangle =  \, \textsf{W}^{(L)}(\lambda) + \frac{\textsf{W}^{(NL)}(\lambda)}{N^2} + O(N^{-4})\, .   
\end{align}
Given that, as demonstrated in \cite{Billo:2021rdb}, $\langle A_k^+ \rangle_0 = \sqrt{2}\,t_k$, and that in the large $N$ limit the function $t_{2k}$ admits the following expansion \cite{Billo:2023kak} 
\begin{align}
\label{eventraces}
t_{2k} \, = \, \frac{N^{k+1}}{2^k} \frac{(2k)!}{k!(k+1)!} - \frac{N^{k-1}}{2^{k+1}} \frac{(2k)!}{k!(k-1)!} \left(1-\frac{k-1}{6} \right) + O(N^{k-3})\, ,
\end{align} 
we conclude that the planar term $\textsf{W}^{(L)}(\lambda)$ is entirely determined by the large $N$ expansion of $\langle W_+ \rangle_0$ alone. In contrast, the next-to-planar term $\textsf{W}^{(NL)}(\lambda)$ receives an additional contribution from the non-trivial vacuum expectation value of the one-point function $\langle \mathcal{P}_{2q}^{+} \rangle$ (see \eqref{1pt}) Therefore, by using the expansion \eqref{eventraces} together with the identity \begin{align}
\label{Identity:SumBesselEven1}
\sum_{q=1}^{\infty} (2q) I_{2q}(x) = \frac{x}{2} I_1(x)\,,
\end{align} we can explicitly determine the first two coefficients appearing on the right-hand side of \eqref{vevW0}. After performing some algebraic manipulations, we obtain
\begin{subequations}
\begin{align}
& \textsf{W}^{(L)}(\lambda) = \frac{2\sqrt{2}\,I_1(\sqrt{\lambda})}{\sqrt{\lambda}}\, \ , 
\label{WL}\\
& \textsf{W}^{(NL)}(\lambda) = \frac{\sqrt{2}}{48}\left(\lambda\,I_0(\sqrt{\lambda})-14\sqrt{\lambda}I_1(\sqrt{\lambda})\right) - \frac{\lambda^{3/2}\partial_{\lambda}\mathcal{F}}{2\sqrt{2}}I_1(\sqrt{\lambda})\, \ .
\label{WNL}
\end{align}
\label{WLandWNL}
\end{subequations}
We note that the expressions \eqref{WLandWNL} are exact in $\lambda$, and will be used to derive the large-$N$ expansion of the integrated correlator \eqref{IZ2}.
\subsection{The connected correlator \texorpdfstring{$\langle W_{+}\mathcal{M}_{\mathbb{Z}_2} \rangle - \langle W_{+} \rangle \langle \mathcal{M}_{\mathbb{Z}_2} \rangle$}{}}
\label{subsec:LargeNIZ2}
Let's now move to consider the connected correlator $\langle W_{+}\mathcal{M}_{\mathbb{Z}_2} \rangle - \langle W_{+} \rangle \langle \mathcal{M}_{\mathbb{Z}_2} \rangle$, which appears in the numerator of \eqref{IZ2}. As a first step, using the decomposition  \eqref{Mexpansion}, we rewrite it as the sum of three connected correlators, namely
\begin{align}
\langle W_{+}\, \mathcal{M}_{\mathbb{Z}_2}\rangle-\langle W_{+} \rangle\langle \mathcal{M}_{\mathbb{Z}_2}\rangle = \sum_{j=0}^{2} \left(\langle W_{+}\,\mathcal{M}_{\mathbb{Z}_2}^{(j)}\rangle - \langle W_{+} \rangle \langle \mathcal{M}_{\mathbb{Z}_2}^{(j)} \rangle\right) \, \equiv \, \sum_{j=0}^{2}\langle W_{+}\,\mathcal{M}_{\mathbb{Z}_2}^{(j)}\rangle_{\text{con}}\, \ .
\label{Mascon}
\end{align}
Then, using the expression \eqref{WuPbasis}  along with the large-$N$ expansions \eqref{PPcorrelators}, it is straightforward to verify that $\langle W_{+}\,\mathcal{M}_{\mathbb{Z}_{2}}^{(0)}\rangle_{\text{con}}$ vanishes, while the remaining two connected correlators are given by
\begin{subequations}
\begin{align}
\langle W_{+}\,\mathcal{M}_{\mathbb{Z}_{2}}^{(1)}\rangle_{\text{con}} &=\sqrt{2}\sum_{q=1}^{\infty}\sqrt{2q}\,I_{2q}(\sqrt{\lambda})\textsf{M}_{0,2q} +\frac{1}{N^2}\left[\frac{\lambda}{32\pi^2}\sum_{q=1}^{\infty}\sqrt{2q}\,I_{2q}(\sqrt{\lambda})\left(\sqrt{2}\,\textsf{Q}_{0,2q}-\frac{\textsf{Q}_{2,2q}}{6}\right) \right. \nonumber \\
& \left. + \sqrt{2}\sum_{k=1}^{\infty}\sum_{q=1}^{\infty}\textsf{M}_{0,2k}\,\textsf{T}_{k,q}^{+}\,\sqrt{2q}\,I_{2q}(\sqrt{\lambda})   -(\lambda\partial_{\lambda}\mathcal{F})^2\sum_{k=1}^{\infty}\sqrt{k}\,\textsf{M}_{0,2k}\sum_{q=1}^{\infty}(2q)I_{2q}(\sqrt{\lambda}) \right]\nonumber \\[0.4em]
&+O(N^{-4})\, \ , \label{M1con}\\[1em]
\langle W_{+}\,\mathcal{M}_{\mathbb{Z}_{2}}^{(2)}\rangle_{\text{con}} & = \frac{1}{2\sqrt{2}N^2}\sum_{\ell=1}^{\infty}(2\ell)\,I_{2\ell}(\sqrt{\lambda})\sum_{q,p=2}^{\infty}(-1)^{p-p\,q}\,\textsf{M}_{p,q}(\sqrt{p\,q}-\textsf{d}_p\,\textsf{d}_q) \nonumber \\
& -\frac{\lambda\partial_{\lambda}\mathcal{F}}{N^2}\sum_{q,p=1}^{\infty}\,\sqrt{p}\,\textsf{M}_{2p,2q}\,\sqrt{2q}\,I_{2q}(\sqrt{\lambda})
 + O(N^{-4})\, \ . 
\label{M2con}
\end{align}
\label{WMcon}
\end{subequations}
It is convenient to express the large-$N$ expansion of the connected correlator \eqref{Mascon} as
\begin{align}
\langle W_{+}\, \mathcal{M}_{\mathbb{Z}_2}\rangle-\langle W_{+} \rangle\langle \mathcal{M}_{\mathbb{Z}_2}\rangle \, = \, \langle W_{+}\mathcal{M}_{\mathbb{Z}_2} \rangle_{\text{con}}^{(L)} + \frac{\langle W_{+}\mathcal{M}_{\mathbb{Z}_2} \rangle_{\text{con}}^{(NL)}}{N^2} + O(N^{-4})\, \ ,   
\label{WMconLargeN}
\end{align}
where the planar term \( \langle W_{+}\mathcal{M}_{\mathbb{Z}_2} \rangle_{\text{con}}^{(L)} \) and the next-to-planar term \( \langle W_{+}\mathcal{M}_{\mathbb{Z}_2} \rangle_{\text{con}}^{(NL)} \) can be easily determined using the expressions \eqref{WMcon}. We finally notice that substituting the expansions \eqref{WMcon} and \eqref{vevW0} into \eqref{IZ2} we obtain exact expressions, valid for any value of the 't Hooft coupling, for the planar term and next-to-planar term of the large $N$ expansion of the integrated correlator  $\mathcal{I}_{\mathbb{Z}_2}$. In the following, we will analyze them  separately, along with their behaviors at strong coupling.
\subsection{The planar term \texorpdfstring{$\mathcal{I}_{\mathbb{Z}_2}^{(L)}$}{}}
Upon examining the expressions \eqref{WMconLargeN}, \eqref{WMcon} and \eqref{WLandWNL}, it becomes clear that the planar term of the large-$N$ expansion  of the integrated correlator \eqref{IZ2} is given by
\begin{align}
\mathcal{I}_{\mathbb{Z}_2}^{(L)} & = \frac{\langle W_{+}\mathcal{M}_{\mathbb{Z}_2} \rangle_{\text{con}}^{(L)}}{\textsf{W}^{(L)}(\lambda)}  \nonumber \\
& = \frac{2}{\textsf{W}^{(L)} (\lambda)}\left(\sqrt{2}\sum_{q=1}^{\infty}\sqrt{2q}\,I_{2q}(\sqrt{\lambda})\textsf{M}_{0,2q}\right) = \frac{\sqrt{\lambda}}{I_1(\sqrt{\lambda})}\sum_{q=1}^{\infty}\sqrt{2q}I_{2q}(\sqrt{\lambda})\textsf{M}_{0,2q}\, \ .
\label{IZ2PlanarStart}
\end{align}
Then, using the explicit expression for the coefficients $\textsf{M}_{0,2q}$ (see \eqref{M0n}), along with the identity
\begin{align}
\sum_{q=1}^{\infty}(-1)^q(2q)I_{2q}(\sqrt{\lambda})J_{2q}(x)  = -\frac{\sqrt{\lambda}\,x}{2(x^2+\lambda)}\left(\sqrt{\lambda}\,I_0(\sqrt{\lambda})\,J_1(x)-x\,I_1(\sqrt{\lambda})\,J_0(x)\right)
\label{sumq}
\end{align}
we can rewrite \eqref{IZ2PlanarStart} as \begin{align} 
\frac{2\pi\sqrt{\lambda}}{I_1(\sqrt{\lambda})}\int_0^{\infty}\frac{dt}{t}\frac{(t/2)^2}{\sinh(t/2)^2}\,\frac{1}{4\pi^2+t^2}\,J_1\left(\frac{t\sqrt{\lambda}}{2\pi}\right)\left[2\pi\,I_0(\sqrt{\lambda})J_1\left(\frac{t\sqrt{\lambda}}{2\pi}\right)-tI_1(\sqrt{\lambda})J_0\left(\frac{t\sqrt{\lambda}}{2\pi}\right)\right]\, \ .
\label{IZ2Planar}
\end{align}
This is our final expression for the planar term of the integrated correlator $\mathcal{I}_{\mathbb{Z}_2}$. Finally, by utilizing the properties of Bessel functions and following the procedure outlined in Appendix \ref{app:StrongCouplingExpansions}, we can easily evaluate the first terms of its expansion at strong coupling, which are given by
\begin{align}
\mathcal{I}_{\mathbb{Z}_2}^{(L)} \, \underset{\lambda \rightarrow \infty}{\sim} \, \frac{\sqrt{\lambda}}{2} +\left(\frac{1}{4}-\frac{\pi^2}{6}\right) + O(\lambda^{-1/2})\, \ .
\label{IZ2PlanarStrong}
\end{align}
We observe that the expressions \eqref{IZ2Planar} and \eqref{IZ2PlanarStrong} coincide with the planar term of the corresponding $\mathcal{N}=4$ SYM integrated correlator \cite{Pufu:2023vwo, Dempsey:2024vkf, Billo:2023ncz, Billo:2024kri}. This agreement is expected, as the integrated correlator \eqref{IZ2} is an untwisted observable. To identify the first deviation from the maximally supersymmetric theory, in the following section, we perform the computation of the next-to-planar term.

\subsection{The next-to-planar term \texorpdfstring{$\mathcal{I}_{\mathbb{Z}_2}^{(NL)}$}{}}
By employing the large $N$-expansions \eqref{vevW0} and \eqref{WMconLargeN}, we find that the next-to-planar term $\mathcal{I}_{\mathbb{Z}_2}^{(NL)}$ of the connected correlator \eqref{IZ2} takes the following form
\begin{align}
\mathcal{I}_{\mathbb{Z}_2}^{(NL)} = \frac{\langle W_{+}\mathcal{M}_{\mathbb{Z}_2} \rangle_{\text{con}}^{(NL)}}{\textsf{W}^{(L)}} - \frac{\langle W_{+}\mathcal{M}_{\mathbb{Z}_2} \rangle_{\text{con}}^{(L)}\,\textsf{W}^{(NL)}}{(\textsf{W}^{(L)})^2}\, \ .
\label{IZ2NL}
\end{align}
Then, using the expressions \eqref{WLandWNL} and \eqref{WMcon},
 we can easily derive its perturbative expansion, whose first terms read
\begin{align}
\mathcal{I}_{\mathbb{Z}_2}^{(NL)}  =   -\frac{3\,\zeta_3}{64\pi^2}\lambda^2 + \left(\pi^2\zeta_3 + 55\,\zeta_5\right)\frac{\lambda^3}{512\pi^4} -\left(27\zeta_3^2+\frac{2625\,\zeta_7}{8}+\frac{25\pi^2\,\zeta_5}{4}+\frac{\pi^4\zeta_3}{4}\right)\frac{\lambda^4}{2048\pi^6} +O(\lambda^5)\, \ . 
\end{align}


In the following, we focus exclusively on deriving the strong coupling expansion of $\mathcal{I}_{\mathbb{Z}_2}^{(NL)}$. In this regard, it is useful to analyze separately the large-$\lambda$ behaviour of the two connected correlators \eqref{M1con} and \eqref{M2con}.
\subsubsection{Strong coupling evaluation of \texorpdfstring{$\langle W_{+}\,\mathcal{M}_{\mathbb{Z}_{2}}^{(1)}\rangle_{\text{con}}^{(NL)}(\textsf{W}^{(L)})^{-1}$}{}}
The next-to-planar contribution of the connected correlator 
\eqref{M1con} reads
\begin{align}
& \frac{1}{\sqrt{2}}\sum_{q=1}^{\infty}\sum_{k=1}^{\infty}(2q)I_{2q}(\sqrt{\lambda})\textsf{M}_{0,2k}\sqrt{2k}\, \left[\frac{(q^2+k^2-1)(q^2+k^2-14)}{12}-(q^2+k^2-1)\lambda\partial_{\lambda}\mathcal{F}-(\lambda\partial_{\lambda})^2\mathcal{F})\right]\nonumber \nonumber \\
& + \frac{\lambda}{16\pi^2\sqrt{2}}\sum_{q=1}^{\infty}\sqrt{2q}I_{2q}\left(\textsf{Q}_{0,2q}-\frac{\textsf{Q}_{2,2q}}{6\sqrt{2}}\right) \, \ .
\label{startM1con}
\end{align}
Let us begin by considering the first line of \eqref{startM1con}. Expanding the terms inside the brackets yields a fourth-degree polynomial in \(q\) and \(k\). We then analyze each distinct monomial separately. In particular, the sums over the monomials of degree zero reads
\begin{align}
\frac{1}{\sqrt{2}}\sum_{q=1}^{\infty}\sum_{k=1}^{\infty}(2q)I_{2q}(\sqrt{\lambda})\sqrt{2k}\,\textsf{M}_{0,2k}\left[\frac{7}{6}+\lambda\partial_{\lambda}\mathcal{F}-(\lambda\partial_{\lambda})^2\mathcal{F}\right]\,  .
\label{StartGradoZero}
\end{align}
Specifically, we observe that the sums over $q$ can be performed using the relation \eqref{Identity:SumBesselEven1}, while the sums over $k$ can be done by exploiting the following identity \cite{Billo:2023kak}
\begin{align}
& \sum_{k=1}^{\infty}\textsf{M}_{0,2k}\sqrt{2k} = -\textsf{M}_{1,1}\, \, .
\label{M02kidentity}
\end{align}
Next, we divide the resulting expression by \eqref{WL}, thereby obtaining the contribution of the degree-zero monomials, namely
\begin{align}
\label{gradoZERO}
-\frac{\sqrt{\lambda}\,\textsf{M}_{1,1}\,I_1(\sqrt{\lambda})}{2\sqrt{2}\,\textsf{W}^{(L)}}\left[\frac{7}{6}+\lambda\partial_{\lambda}\mathcal{F}-(\lambda\partial_{\lambda})^2\mathcal{F}\right] = -\frac{\lambda\,\textsf{M}_{1,1}}{8}\left[\frac{7}{6}+\lambda\partial_{\lambda}\mathcal{F}-(\lambda\partial_{\lambda})^2\mathcal{F}\right]\, \ .
\end{align}

The summation over the monomials of degree two can be performed using the expressions \eqref{Identity:SumBesselEven1}, \eqref{M02kidentity}, \eqref{id2} and the following identity\footnote{This identity was demonstrated in Appendix C of \cite{Pini:2024uia}.}
\begin{align}
\sum_{k=1}^{\infty}\sqrt{k}\,k^2\,\textsf{M}_{0,2k} = -\frac{1}{2\sqrt{2}}(1+\lambda\partial_{\lambda})\textsf{M}_{1,1}\, \ .   
\end{align}
Therefore, after some algebra and dividing by \eqref{WL}, we find that the contribution due to  monomials of degree two is given by
\begin{align}
\frac{\lambda}{16}\left[\frac{5}{4}+\lambda\partial_{\lambda}\mathcal{F}\right]\left(1+\frac{I_0(\sqrt{\lambda})}{I_1(\sqrt{\lambda})}\sqrt{\lambda}+\lambda\partial_{\lambda}\right)\textsf{M}_{1,1}\, \ .
\label{gradoDUE}
\end{align}
Finally, the summation over the monomials of degree four can be carried out using the additional identity \eqref{id3}. After some algebra and taking the ratio with \eqref{WL} then yields
\begin{align}
\frac{\lambda}{192}\left[2\,\mathcal{B}(\lambda)-\sqrt{\lambda}\left(\sqrt{\lambda}+2\frac{I_0(\sqrt{\lambda})}{I_1(\sqrt{\lambda})}+\frac{I_0(\sqrt{\lambda})}{I_1(\sqrt{\lambda})}\lambda\partial_{\lambda}\right)\textsf{M}_{1,1}\right]\, ,
\label{gradoQUATTRO}
\end{align}
where we introduced the function\footnote{We refer the reader to Appendix \ref{app:Identities} for the  derivation of the r.h.s. of this expression.}
\begin{align}
\mathcal{B}(\lambda)  & \equiv \sum_{k=1}^{\infty}\sqrt{2k}\,k^4\,\textsf{M}_{0,2k}\nonumber \\  
& = \frac{\sqrt{\lambda}}{8\pi^2}\int_0^{\infty}dt\,\frac{(t/2)^2}{\sinh(t/2)^2}\,J_1\left(\frac{t\sqrt{\lambda}}{2\pi}\right)\left[2\pi\,J_0\left(\frac{t\sqrt{\lambda}}{2\pi}\right)-t\sqrt{\lambda}J_1\left(\frac{t\sqrt{\lambda}}{2\pi}\right)\right]\, \ .
\label{BB}
\end{align}
All the expressions that we got so far are valid for any value of the 't Hooft coupling. Let's now consider their the strong coupling limit, i.e. $\lambda >>1$. First of all we recall that  the strong coupling expansion of the free energy $\mathcal{F}(\lambda)$ of the $\mathbb{Z}_2$ quiver gauge theory was performed in \cite{Beccaria:2022ypy}, where it was found that
\begin{align}
\mathcal{F}(\lambda) \, \underset{\lambda \rightarrow \infty}{\sim} \, \frac{\sqrt{\lambda}}{4}-\frac{1}{2}\log \lambda + O(\lambda^0)\, \ .
\label{Fstrong}
\end{align}
Moreover, the strong coupling expansion of the matrix elements $\textsf{M}_{n,m}$ is given by \eqref{MnmStrong}. Then, by exploiting the asymptotic expansion of the modified Bessel functions $I_n(\sqrt{\lambda})$, we evaluate the strong coupling behaviour of \eqref{gradoZERO} and \eqref{gradoDUE}. We obtain
\begin{subequations}
\begin{align}
-\frac{\textsf{M}_{1,1}\,\lambda}{8}\left[\frac{7}{6}+\lambda\partial_{\lambda}\mathcal{F}-(\lambda\partial_{\lambda})^2\mathcal{F}\right] & \, \underset{\lambda \rightarrow \infty}{\sim}\,  \frac{\lambda^{3/2}}{256}+\frac{13\,\lambda}{384} + O(\lambda^{1/2})\, \ , \\[0.5em]
\frac{\lambda}{16}\left(\frac{5}{4}+\lambda\partial_{\lambda}\mathcal{F}\right)\left[1+\frac{\sqrt{\lambda}\,I_0(\sqrt{\lambda})}{I_1(\sqrt{\lambda})}+\lambda\partial_{\lambda}\right]\textsf{M}_{1,1} & \, \underset{\lambda \rightarrow \infty}{\sim} \, -\frac{\lambda^2}{256}-\frac{11\lambda^{3/2}}{512} + O(\lambda)\, \ .
\end{align}
\label{grado0e2StrongCoupling}
\end{subequations}
Finally, to derive the strong coupling expansion of \eqref{gradoQUATTRO}, we must evaluate the large $\lambda$ behavior of \eqref{BB}. This calculation is carried out in Appendix \ref{app:StrongCouplingExpansions}, resulting in the expression \eqref{BBstrong}. Thus, using such result, we obtain
\begin{align}
\frac{\lambda}{192}\left[2\,\mathcal{B}(\lambda)-\sqrt{\lambda}\left(\sqrt{\lambda}+2\frac{I_0(\sqrt{\lambda})}{I_1(\sqrt{\lambda})}+\frac{I_0(\sqrt{\lambda})}{I_1(\sqrt{\lambda})}\lambda\partial_{\lambda}\right)\textsf{M}_{1,1}\right] \, \underset{\lambda \rightarrow \infty}{\sim} \,  \frac{\lambda^2}{384}  -\frac{\lambda^{3/2}}{1152} + O(\lambda)\, \ . 
\label{grado4StrongCoupling}
\end{align}
The two contributions in the second line of \eqref{startM1con} can be evaluated following the same procedure used for the planar term. Specifically, we use the definition \eqref{Qmatrix} for the coefficients $\textsf{Q}_{0,2q}$ and $\textsf{Q}_{2,2q}$ and perform the sums over $q$ using the identity \eqref{sumq}. Subsequently, by exploiting the asymptotic properties of the Bessel functions, we determine the strong coupling expansions
\begin{subequations}
\begin{align}
& \frac{\lambda}{16\pi^2\sqrt{2}\,\textsf{W}^{(L)}}\sum_{q=1}^{\infty}\sqrt{2q}\,I_{2q}(\sqrt{\lambda})\textsf{Q}_{0,2q} \, \underset{\lambda \rightarrow \infty}{\sim} \,  \frac{\lambda}{96}\left(\pi^2-9\right) + O(\lambda^{1/2})\, , \\
& \frac{\lambda}{192\pi^2\,\textsf{W}^{(L)}}\sum_{q=1}^{\infty}\sqrt{2q}\,I_{2q}(\sqrt{\lambda})\textsf{Q}_{2,2q} \,\underset{\lambda \rightarrow \infty}{\sim} \, \frac{\lambda^{3/2}}{1152}\left(\pi^2-10\right)+ O(\lambda)\, \ .
\end{align}
\label{QtermsStrong}
\end{subequations}

As a final step, we sum the expressions \eqref{grado0e2StrongCoupling}, \eqref{grado4StrongCoupling} and \eqref{QtermsStrong}. This allows to determine the leading terms of the strong coupling expansion
\begin{align}
\frac{\langle W_{+}\,\mathcal{M}_{\mathbb{Z}_{2}}^{(1)}\rangle_{\text{con}}^{(NL)}}{\textsf{W}^{(L)}} \, \underset{\lambda \rightarrow \infty}{\sim} \, -\frac{\lambda ^2}{768} -\frac{\pi ^2 \lambda ^{3/2}}{1152}-\frac{5 \lambda ^{3/2}}{512}+O(\lambda)\, \ .
\label{con1STRONG}
\end{align}
\subsubsection{Strong coupling evaluation of \texorpdfstring{$\langle W_{+}\,\mathcal{M}_{\mathbb{Z}_{2}}^{(2)}\rangle_{\text{con}}^{(NL)}(\textsf{W}^{(L)})^{-1}$}{}}
Let us now consider the second connected correlator and begin by analyzing the term in the second line of \eqref{M2con}. We perform the sums over $q$ and $p$ using the explicit expression of the coefficients $\textsf{M}_{n,m}$ (given in \eqref{Mnm}), the relation \eqref{sumq} and the following identity
\begin{align}
\sum_{p=1}^{\infty}(-1)^{p}(2p)J_{2p}(x) = -\frac{x}{2}J_1(x)\, .   
\end{align}
After some algebraic manipulations we obtain
\begin{align}
& \sum_{p,q=1}^{\infty}\sqrt{p}\,\textsf{M}_{2p,2q}\,\sqrt{2q}\,I_{2q}(\sqrt{\lambda}) =\nonumber \\
& -\frac{\lambda}{4\sqrt{2}}\int_{0}^{\infty}dq\,\frac{q^3}{\sinh(q)^2}\,\frac{1}{\pi^2+q^2}\,J_1\left(\frac{q\sqrt{\lambda}}{\pi}\right)\left[I_0(\sqrt{\lambda})J_1\left(\frac{q\sqrt{\lambda}}{\pi}\right)-\frac{q}{\pi}I_1(\sqrt{\lambda})J_0\left(\frac{q\sqrt{\lambda}}{\pi}\right)\right]\, \ .   
\end{align}
Next, we divide by \eqref{WL} and derive the strong coupling expansion of the resulting expression by exploiting the asymptotic behavior of the Bessel functions and using the large $\lambda$ expansion of the free energy \eqref{Fstrong}. This results in
\begin{align}
-\frac{\lambda\partial_{\lambda}\mathcal{F}}{\textsf{W}^{(L)}}\sum_{p,q=1}^{\infty}\sqrt{p}\,\textsf{M}_{2p,2q}\,\sqrt{2q}\,I_{2q}(\sqrt{\lambda}) \, \underset{\lambda \rightarrow \infty}{\sim} \,    \frac{\lambda^{3/2}}{768}\left(\pi^2-9\right) + O(\lambda)\, \ .
\label{MIstrong}
\end{align}
The terms in the first line of \eqref{M2con} require more attention. First, we observe that the sums over $\ell$ and the sums over $q$ and $p$ factorize. The sums over $\ell$ can be performed analytically using the identity \eqref{id5}, while the leading term of the strong coupling expansion of the sums over $q$ and $p$ was numerically evaluated in \cite{DeSmet:2025mbc,Pini:2024uia}, where it was found that\footnote{The evaluation carried out in \cite{Pini:2024uia}, based on a Padé approximant, turned out to be less accurate, as it led to the estimate $-\frac{\pi^2}{240}$.}
\begin{align}
\sum_{p,q=2}^{\infty}(-1)^{p-pq}\,\textsf{M}_{p,q}\,(\sqrt{p\,q}-\textsf{d}_p\,\textsf{d}_q)\, \underset{\lambda \rightarrow \infty}{\sim}\, -\frac{1}{24}\sqrt{\lambda} + O(\lambda^0)\, \ .
\end{align}
Therefore, after dividing by \eqref{WL}, we obtain the following strong coupling expansion
\begin{align}
\frac{1}{2\sqrt{2}\,\textsf{W}^{(L)}}\sum_{\ell=1}^{\infty}(2\ell)\,I_{2\ell}(\sqrt{\lambda})\sum_{q,p=2}^{\infty}(-1)^{p-pq}\textsf{M}_{p,q}(\sqrt{pq}-\textsf{d}_p\textsf{d}_q)   \underset{\lambda \rightarrow \infty}{\sim} -\frac{\lambda ^{3/2}}{384} + O(\lambda)\, \ . 
\label{M2strong}
\end{align}
Finally, we sum up the expressions \eqref{MIstrong} and \eqref{M2strong}, this yields
\begin{align}
\frac{\langle W_{+}\,\mathcal{M}_{\mathbb{Z}_{2}}^{(2)}\rangle_{\text{con}}^{(NL)}}{\textsf{W}^{(L)}}  \, \underset{\lambda \rightarrow \infty}{\sim} \, \frac{\lambda^{3/2}}{768} \left(\pi ^2-11\right)  + O(\lambda)\, \ .
\label{con2STRONG}
\end{align}

\subsubsection*{Strong coupling expansion of  \texorpdfstring{$\mathcal{I}_{\mathbb{Z}_2}^{(NL)}$}{}}
Finally, by using the relation \eqref{IZ2NL} as well as the expressions \eqref{con1STRONG}, \eqref{con2STRONG} and \eqref{WLandWNL}, we obtain the strong coupling expansion
\begin{align}
\mathcal{I}^{(NL)}_{\mathbb{Z}_2} \,\underset{\lambda \rightarrow \infty}{\sim} \, - \frac{\lambda^{3/2}}{128} + O(\lambda)\, \ .
\end{align}
This our final result for the large $\lambda$ expansion of this observable.

\section{Conclusions}
\label{sec:Conclusions}
In this work, by exploiting supersymmetric localization, we studied the large $N$ limit of the integrated correlator \eqref{Wint} between a Wilson line and two local Higgs branch operators of conformal dimension 2. Specifically, we obtained exact expressions for the planar and next-to-planar terms of its large $N$ expansion, valid for any value of the 't Hooft coupling $\lambda$. Then, by utilizing the asymptotic properties of the Bessel functions, we derived the first terms of the corresponding large $\lambda$ expansion, which reads
\begin{align}
\mathcal{I}_{\mathbb{Z}_2} \, \underset{\lambda \rightarrow \infty}{\sim} \, \frac{\sqrt{\lambda}}{2} +\left(\frac{1}{4}-\frac{\pi^2}{6}\right) + O\left(\frac{1}{\sqrt{\lambda}}\right)\,  + \frac{1}{N^2}\left[-\frac{\lambda^{3/2}}{128}  + O(\lambda)\right] + O(N^{-4})\, \ .
\label{IZ2StrongSummary}
\end{align}

In the future, it would be interesting to extend the set of $\mathcal{N}=2$ gauge theories for which the large $N$ computation of the integrated correlator with the insertion of a defect operator can be performed analytically using supersymmetric localization. A natural candidate in this context would be the $M > 2$ node quiver gauge theory discussed in \cite{Pini:2017ouj}. It is important to note, as mentioned in \cite{Pini:2023lyo}, that the half-BPS Wilson loop associated with any node of the quiver can always be expressed as a linear combination of one untwisted Wilson loop and $M-1$ twisted Wilson loops. However, the integrated correlator \eqref{Wint} with the insertion of a twisted Wilson loop vanishes, leaving us only with possibility represented by the insertion of the untwisted Wilson loop. Furthermore, let $m_I$ (with $ I=0,\ldots,M-1$) denote the masses assigned to the $\mathcal{N}=2$ hypermultiplet in the mass deformed theory. It can be easily shown that terms proportional to $m_I\,m_J$ with $I \neq J$ do not appear in the small mass expansion of the partition function. This, in turn, implies that the only way to obtain a non-vanishing correlator is to take two derivatives with respect to the same mass. Consequently, up to an overall numerical factor (related to the number of nodes of the quiver), we expect that the result for the $M$-nodes quiver must be identical to the one discussed in this article for the $M=2$ case. Nevertheless, it would be intriguing to consider $\mathcal{N}=2$ SCFTs with a different type of interaction action, that includes single-trace contributions. In this context, we plan to consider the so-called ``type-D" theory \cite{Billo:2024ftq} and the $Sp(N)$ gauge theory discussed in \cite{Beccaria:2022kxy}. Work
along these lines is in progress.  
 
Finally, as mentioned in the Introduction, the $\mathcal{N}=2$ theory we considered has a known gravity dual geometry, therefore it would be also interesting to analyse the holographic implication of our result along the lines of \cite{Pufu:2023vwo}.

\vskip 2cm
\noindent {\large {\bf Acknowledgments}}
\vskip 0.2cm
I'm very grateful to Paolo Vallarino for many important discussions and for reading and commenting on the draft of the article. Furthermore, I would like to thank Marco Billò, Lorenzo De Lillo, Pieter-Jan De Smet, Marialuisa Frau and Alberto Lerda for useful discussions. 
The work of AP is supported  by the Deutsche Forschungsgemeinschaft (DFG, German Research Foundation) via the Research Grant ``AdS/CFT beyond the classical supergravity paradigm: Strongly coupled gauge theories and black holes” (project number 511311749). 
\vskip 1cm

\appendix

\section{Mathematical identities and the function \texorpdfstring{$\mathcal{B}(\lambda)$}{}}
\label{app:Identities}
In this appendix, we compile and prove the mathematical identities used in Section~\ref{sec:LargeNIntegartedCorrelator} to derive the large $N$ expansion of the integrated correlator $\mathcal{I}_{\mathbb{Z}_2}$.
\begin{itemize}
\item Proof of
\begin{align}
\sum_{q=2}^{\infty}qI_{q}(x) = \frac{x}{2}\left(I_1(x)+I_2(x)\right)\, \ .
\label{id5}
\end{align}
We separate the sums into contributions from even and odd values of $q$ then, using the definition of $I_q(x)$, we obtain
\begin{align}
& \sum_{q=2}^{\infty}q\,I_{q}(x) = \sum_{q=1}^{\infty}(2q)I_{2q}(x)+\sum_{q=1}^{\infty}(2q+1)I_{2q+1}(x) = \sum_{q=1}^{\infty}\sum_{m=0}^{\infty}\frac{2q}{m!(m+2q)!}\left(\frac{x}{2}\right)^{2m+2q} \nonumber \\
& + \sum_{q=1}^{\infty}\sum_{m=0}^{\infty}\frac{2q+1}{m!(m+2q+1)!}\left(\frac{x}{2}\right)^{2m+2q+1} = \sum_{n=1}^{\infty}\sum_{q=1}^{n}\frac{2q}{(n-q)!(n+q)!}\left(\frac{x}{2}\right)^{2n} \nonumber \\
& + \sum_{n=1}^{\infty}\sum_{q=1}^{n}\frac{2q+1}{(n-q)!(n+q+1)!}\left(\frac{x}{2}\right)^{2n+1} = \sum_{k=0}^{\infty}\frac{1}{k!(k+1)!}\left(\frac{x}{2}\right)^{2k+2}+ \sum_{k=0}^{\infty}\frac{1}{k!(k+2)!}\left(\frac{x}{2}\right)^{2k+3} \nonumber\\
& = \frac{x}{2}\left(I_1(x)+I_2(x)\right)\, \ .
\end{align}
\item Proof of
\begin{align}
& 2\sum_{q=1}^{\infty}q^3I_{2q}(x) = \frac{x^2}{4}I_0(x)
\label{id2} 
\end{align}
Using the definition of $I_{2q}(x)$, we obtain
\begin{align}
& 2\sum_{q=1}^{\infty}q^3I_{2q}(x) = 2\sum_{q=1}^{\infty}\sum_{m=0}^{\infty}\frac{q^3}{m!(m+2q)!}\left(\frac{x}{2}\right)^{2m+2q} = 2\sum_{n=1}^{\infty}\left(\frac{x}{2}\right)^{\frac{n}{2}}\sum_{q=1}^{n}\frac{q^3}{(n-q)!(n+q)!} = \nonumber \\
& \sum_{n=1}^{\infty}\frac{1}{(n-1)!(n-1)!}\left(\frac{x}{2}\right)^{2n} = \sum_{k=0}^{\infty}\frac{1}{k!\,k!}\left(\frac{x}{2}\right)^{2k+2}=\frac{x^2}{4}\sum_{k=0}^{\infty}\frac{1}{k!\,k!}\left(\frac{x}{2}\right)^{2k} = \frac{x^2}{4}I_0(x)\,\ . 
\end{align}

\item Proof of
\begin{subequations}
\begin{align}
& 2\sum_{k=1}^{\infty}(-1)^k\,k^5\,J_{2k}(x) = \frac{x^2}{4}\left(x\,J_1(x)-J_0(x)\right)\, \ , \label{id30}\\
& 2\sum_{q=1}^{\infty}q^5I_{2q}(x)   =  \frac{x^2}{4}\left(x\,I_1(x)+I_0(x)\right)\, \ .
\label{id3}    
\end{align}
\label{idIJ5}
\end{subequations}
Let us prove \eqref{id3}, as the demonstration of \eqref{idIJ5} is completely analogous. Using the definition of $I_{2q}(x)$, we obtain
\begin{align}
& 2\sum_{q=1}^{\infty}q^5I_{2q}(x) = 2\sum_{q=1}^{\infty}\sum_{m=0}^{\infty}\frac{q^5}{m!(m+2q)!}\left(\frac{x}{2}\right)^{2m+2q} = 2\sum_{n=1}^{\infty}\left(\frac{x}{2}\right)^{\frac{n}{2}}\sum_{q=1}^{n}\frac{q^5}{(n-q)!(n+q)!} = \nonumber\\
& \sum_{n=1}^{\infty}\frac{2n-1}{(n-1)!(n-1)!}\left(\frac{x}{2}\right)^{2n} =  2\sum_{m=0}^{\infty}\frac{(m+1)}{m!\,m!}\left(\frac{x}{2}\right)^{2m+2}  -\frac{x^2}{4}I_0(x) = 2\sum_{k=0}^{\infty}\frac{1}{k!(k+1)!}\left(\frac{x}{2}\right)^{2k+4} \nonumber \\
& +\frac{x^2}{4}I_0(x) = \frac{x^2}{4}\left(x\,I_1(x)+I_0(x)\right)\, \ .
\end{align}
\item Proof of
\begin{align}
\mathcal{B}(\lambda) =
& \frac{\sqrt{\lambda}}{8\pi^2}\int_0^{\infty}dt\,\frac{(t/2)^2}{\sinh(t/2)^2}J_1\left(\frac{t\sqrt{\lambda}}{2\pi}\right)\left[2\pi\,J_0\left(\frac{t\sqrt{\lambda}}{2\pi}\right)-t\sqrt{\lambda}J_1\left(\frac{t\sqrt{\lambda}}{2\pi}\right)\right]
\label{id4}
\end{align}
Using the definition of the $\textsf{M}_{0,2k}$ coefficients given in \eqref{M0n}, we obtain
\begin{align}
\sum_{k=1}^{\infty}\sqrt{2k}\,k^4\,\textsf{M}_{0,2k} = - \int_0^{\infty}\frac{dt}{t}\frac{(t/2)^2}{\sinh(t/2)^2}\left(\frac{8\pi}{t\sqrt{\lambda}}\right)J_1\left(\frac{t\sqrt{\lambda}}{2\pi}\right)\sum_{k=1}^{\infty}(-1)^k\,k^5\,J_{2k}\left(\frac{t\sqrt{\lambda}}{2\pi}\right)\, .   
\end{align}
We finally perform the sum over $k$ using the identity \eqref{id30} and we finally find
\begin{align}
\frac{\sqrt{\lambda}}{8\pi^2}\int_0^{\infty}dt\,\frac{(t/2)^2}{\sinh(t/2)^2}\,J_1\left(\frac{t\sqrt{\lambda}}{2\pi}\right)\left[2\pi\,J_0\left(\frac{t\sqrt{\lambda}}{2\pi}\right)-t\sqrt{\lambda}J_1\left(\frac{t\sqrt{\lambda}}{2\pi}\right)\right]\, \ .    
\end{align}
\end{itemize}

\section{Strong coupling expansions}
\label{app:StrongCouplingExpansions}
In this work, we perform several computations at strong coupling using the Mellin–Barnes method. To illustrate this approach, we apply it to the specific example of the strong coupling evaluation of the expression \eqref{id4}. As a first step we employ the Mellin representation of the product of two Bessel functions, namely
\begin{align}
J_{n}(x)J_{m}(x) = \int_{-i\infty}^{+i\infty} \frac{ds}{2\pi\,i}\, \frac{\Gamma(-s)\Gamma(2s+n+m+1)}{\Gamma(s+n+1)\Gamma(s+m+1)\Gamma(s+n+m+1)}\left(\frac{x}{2}\right)^{2s+n+m}\, \ .
\label{MellinBessel}
\end{align}
In this way the expression \eqref{id4} becomes
\begin{align}
\mathcal{B}(\lambda) &= \frac{\sqrt{\lambda}}{2\pi}\int_0^{\infty} dq\, \frac{q^2}{\sinh(q)^2}\int_{-i\infty}^{i\infty}\frac{ds}{2\pi i} \frac{\Gamma(-s)\Gamma(2s+2)}{\Gamma(s+2)^2\Gamma(s+1)}\left(\frac{q\sqrt{\lambda}}{2\pi}\right)^{2s+1} \nonumber \\[0.2em] & -\frac{\lambda}{2\pi^2}\int_0^{\infty} dq\, \frac{q^3}{\sinh(q)^2}\int_{-i\infty}^{i\infty}\frac{ds}{2\pi i} \frac{\Gamma(-s)\Gamma(2s+3)}{\Gamma(s+2)^2\Gamma(s+3)}\left(\frac{q\sqrt{\lambda}}{2\pi}\right)^{2s+2}\, \ .   
\end{align}
Then, the integral over the $q$ variable can be performed by exploiting the identity
\begin{align}
\int_0^{\infty}dq\, \frac{q^{1+2s}}{\sinh^2(q)} = 4^{-s}\,\Gamma(2+2s)\zeta(2s+1)\, \ .
\end{align}
In this way, as shown in \cite{Beccaria:2021hvt,Pufu:2023vwo}, we arrive at an expression whose asymptotic expansion receives contributions from the poles on the negative real axis. By summing the residues at these poles, we obtain
\begin{align}
\mathcal{B}(\lambda) \, \underset{\lambda \rightarrow \infty}{\sim} \,  -\frac{\sqrt{\lambda }}{12}+\frac{1}{4}-\frac{1}{16}\frac{1}{\sqrt{\lambda}} +O(\lambda^{-3/2})\, \ .
\label{BBstrong}
\end{align}
By applying the same procedure, we can also derive the large $\lambda$ expansion of the matrix elements $\textsf{M}_{n,m}$ defined in \eqref{Mnm}. The result  is \cite{Billo:2023kak}
\begin{align}
\textsf{M}_{n,m}   \, \underset{\lambda \rightarrow \infty}{\sim} \, -\frac{1}{2}\delta_{n,m} +\frac{\sqrt{n\,m}}{\sqrt{\lambda}} + O(\lambda^{-3/2})\, \ .
\label{MnmStrong}
\end{align}

\printbibliography

\end{document}